# Upgrade of the Analog Integrator for EAST Device

Y. Wang, *Member, IEEE,* Z.S. Ji, Z.C. Zhang, S. Li, F. Wang, *Member, IEEE,* X.Y. Sun, *Member, IEEE*

*Abstract*–Integrators are fundamental instruments to recover differential signals from magnetic probes in Experimental Advanced Superconducting Tokamak (EAST) experiments. A kind of difference integrator is introduced which has the same structure as the standard difference amplifier. The linear fitting method is used for determining the effective drift slope, then the plasma control system (PCS) use the drift slope to rectify the integration signal in real time. A new integrator controller was developed, which uses an ARM micro-controller and the lightweight IP protocol stack to realize the network control. The tests show that the upgraded integrator works well, its CMRR is high up to 125 dB when the common voltage is 1.5 V, and the processed integration drift is about 200 uVs /1000 s.

## I. Introduction

EAST is aimed at high performance plasma for long pulse up to 1000 s or longer [1, 2], while the magnetic diagnostics (MDs) is one of important elements to realize EAST missions [3]. MDs is composed of a series of magnetic sensors or coils, from which the integrators are required to convert acquired voltages to field strength.

There are more than 600 channels of integrators has been used for EAST since 2008, and most of the integrators work well in the EAST campaigns during these years, but some problems are also exposed. The main purpose of the upgrade of the analog integrator are as follow:
- ➢ To optimize the system structure and improve the maintainability and testability.
- ➢ To improve the common mode rejection ration (CMRR).
- ➢ To reduce the integration drift.

The way to improve the CMRR of the integrator is using the differential input integrator instead of the single-ended input integrator [4-8]. There are some solutions to reduce the integration drift have been designed on different superconducting machines [4-11]. Two main development branches are the analog integrator and the digital integrator, while both of them have advantages and disadvantages [4].

In this article, a kind of the differential input integrator was designed to improve the CMRR, and the linear fitting method was used to reduce the integration drift, and a new integrator controller was designed to upgrade the integrator system, and the performance test and analysis has been demonstrated.

## II. Design

### A. Architecture

The architecture of the new integrator system is shown in Fig.1. The input of the integrator is a magnetic signal or a standard test signal. There are two outputs, one is to the Plasma Control System (PCS), and the other is to the Data Acquisition System (DAS). Before transit the signals to integration results, each path will go through remote control amplifiers and isolators.

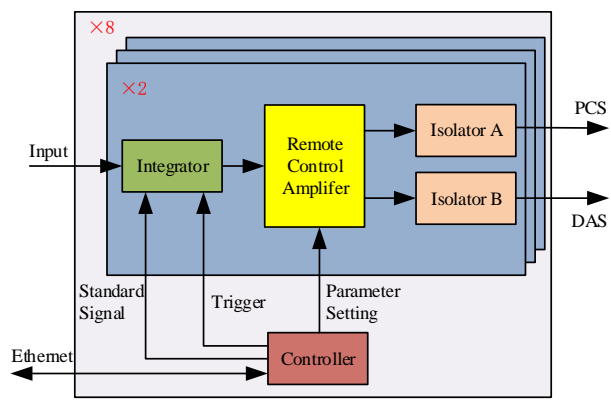

Fig. 1. Architecture of the upgraded integrator.

### B. Controller

The controller of the new integration system has the following functions: Ethernet Transmission Control Protocol/Internet Protocol (TCP/IP) communications, integrator control, gain setting of remote control amplifiers, generating standard signal. Fig.2 is the structure of the controller. The ARM Cortex-M4 CPU STM32F407 is the core of the integrator controller, which has 1 Mbyte of Flash memory and 192 Kbyte of Static Random Access Memory (SRAM) [12]. The external 8 Kbit Electrically Erasable Programmable Read Only Memory (EEPOM) 24C08 [13] is used for parameters storage. STM32F407 achieves 10/100M Ethernet communications by the internal Ethernet MAC controller and an external PHY chip LAN8720A [14]. STM32F407 controls various functions through its input/output port, including the integrator controller, amplifier parameter controller, standard signal generator, and other functions.

Manuscript received June 22, 2018. This work is supported by National Key R&D Program of China (Grant No: 2017YFE0300500, 2017YFE0300504).

Y. Wang is with the Institute of Plasma Physics, Chinese Academy of Sciences, 350 Shushanhu Road, Hefei, Anhui 230031, P. R. China (corresponding author to provide phone: 86-551-5592776; e-mail: wayong@ipp.ac.cn).

Z.S. Ji, Z.C. Zhang, S. Li, F. Wang, X.Y. Sun are with the Institute of Plasma Physics, Chinese Academy of Sciences, 350 Shushanhu Road, Hefei, Anhui 230031, P. R. China.

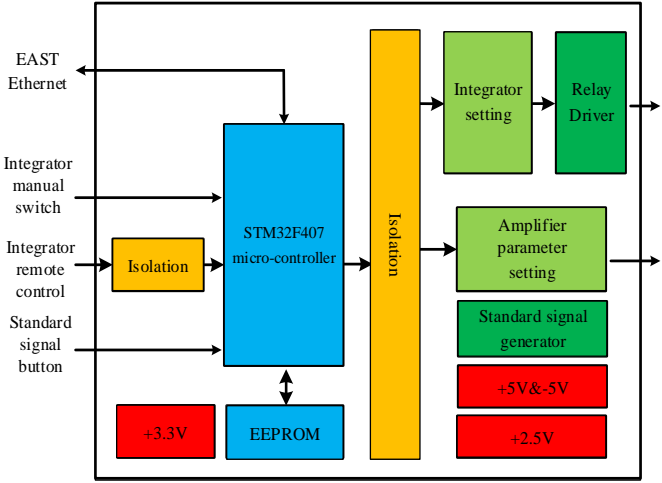

Fig. 2. Schematic of the upgraded controller.

Integrator controller is designed to control the integrator in three ways: the remote control by an external hardware-level signals, local manually switch control and command control based on TCP/IP communications. The first is mainly used in the normal experimental stage. The other two methods are designed for integrator debugging.

Amplifier parameters controller includes the address decoding and output latch circuit. Output latch is connected with the address input of analog switch MAX308 [15] to control the analog switch channels, thereby control the gain of remote control amplifier.

Standard signal generator can provide two types of standard signals. One is a positive pulse of +2.5 V amplitude and 10 ms pulse width, which immediately followed by a negative pulse of -2.5 V amplitude and 10ms pulse width. The other is a positive pulse, whose amplitude is +2.5 V and pulse width is 1, 2, 3 … 9, 10 or 1 s for integrators of different integration time constant. Standard signal is used for checking whether the integrator works well and calibrate the integration time constant of integrators. The reference voltage of 2.5 V is provided by REF5025 [16].

The program on STM32F407 has realized TCP communications based on LwIP, and the instructions for the integrator as shown in Tab. 1.

Tab. 1 Instructions of new controller

| Instruction | Description |
| --- | --- |
| ALLd;d;d;d;d;d;d;d;d | Set the gain for all channels |
| READAll | Read the gain of all channels |
| RCd | Set the gains on one module |
| INTEd | Set same gain for all channels |
| Initialization | Initialize to normal mode |
| StandardSignal | Integration of a standard signal |
| PulseSignal | Integration of a pulse signal (extended to 1s) |
| IntHold | Hold the integration value |
| NET X.X.X.X;X.X.X.X;X.X.X.X | Modify IP, MASK, GATEWAY |
| QUIT | TCP disconnected |

Compared with the original design, the new controller simplifies the design of hardware and software, and improves the system performance, usability, maintainability and testability.

*C. Integrator*

The previous analog integrators are single-ended input, which has low CMRR, a kind of difference integrator is introduced which has the same structure as the standard difference amplifier, in which two resistors are replaced by two capacitors, as shown in Fig.3.

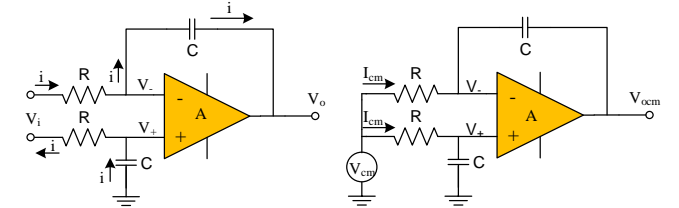

(a) Differential mode input signal    (b) common mode input signal

Fig. 3. Difference integrator.

When the input signal is a differential mode input signal $V_i$, as shown in Fig. 3(a), the output is

$$V_o = -\frac{1}{RC}\int_0^t V_i \mathrm{d}t \quad (1)$$

When the input signal is a common mode input signal $V_{cm}$, as shown in Fig. 3(b), the output is

$$V_{ocm} = 0 \quad (2)$$

The difference integrator has the same integration function of differential mode signal as the single-ended integrator, but it has a much higher rejection of common mode signals. The higher the CMRR is, the better the integrator is. The CMRR of difference integrator may be affected by the resistors and capacitors.

Low noise zero-drift dual operational amplifier LTC1151 [17] is very suitable for integrator, which has unmatched performance by constantly adjusting offset voltage errors through Chopper-stabilization technology. Its typical offset voltage is 0.5 uV, offset current is ±20 pA. Polystyrene capacitors with high leakage resistance and small precision metal film resistors of ±0.5% accuracy and ± 50ppm/°C temperature drift coefficient are used for further reducing the error.

## III. PERFORMANCE

A new set of analog integrator of 20 ms (R = 20 kΩ, C = 1 uF) integration time constant was tested.

*A. Common mode rejection Test*

When the input of the integrator is shorted and the common-mode voltage is 0, the drift is almost 0, as shown in Fig. 4(a). When the common mode voltage is 0.13 V, the drift is also 0, as shown in Fig. 4(b). While when the common mode voltage is 1.5 V, the drift is about 4 mV in 100 s integration, as shown in Fig. 4(c), which means 1.5 V common mode voltage has some affection with the integration drift.

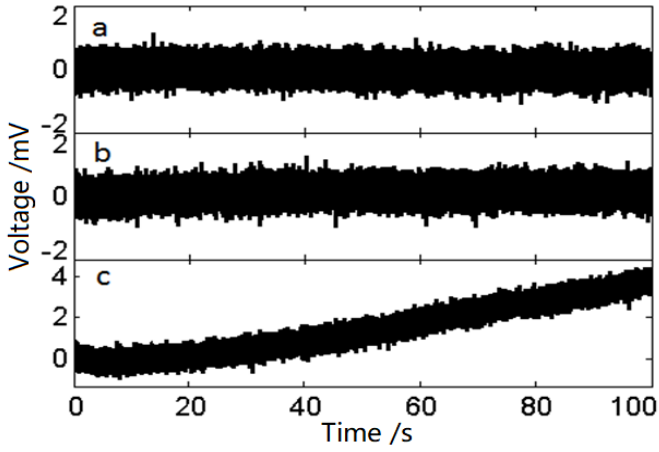

Fig. 4. 100s drifts of 20ms diffence integrator under common mode voltages of 0 V (a), 0.13 V (b) and 1.5 V (c).

Using the definition of the CMRR of the difference amplifier, the CMRR of the difference integrator is

$$CMR = 20\lg \frac{A_{uf}}{A_{cf}} \quad (3)$$
$$= 20\lg \frac{1.5V}{(4mV \times 20ms)/100s} \approx 125 dB$$

Where $A_{uf}$ is the integration drift rate of the difference mode signal, while $A_{cf}$ is the integration drift rate of the common mode signal.

Usually, the common mode signal of mV level may not affect the drift of the integrator.

### B. Drift Test

A reference shot is used for calculating the drift rate, then the plasma control system use the drift slope to rectify the integration signal in the following shot in real time.

The raw signal drift is about 50 mV / 400 s, while the processed signal drift is about 4 mV / 400 s, that is 200 uVs /1000 s (RC=20 ms), as shown in Fig. 5. This performance can satisfy the integration requirements of long pulse plasma discharge on EAST.

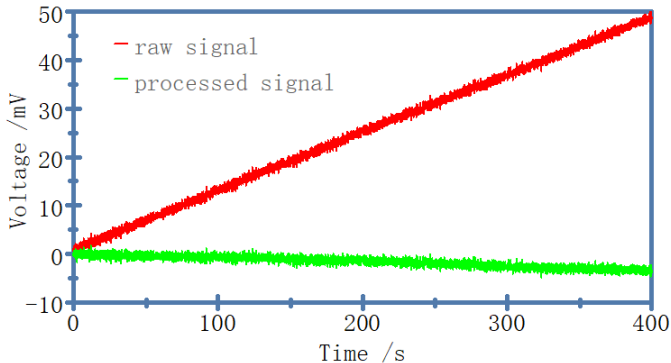

Fig. 5. Integration drifts of raw signal and processed signal.

## IV. CONCLUSION AND FUTURE WORK

This work clearly shows that the analog integrator has been upgrade, which has use the differential input integrator structure to improve the CMRR, and use the reference drift slope to reduce the drift in real time, and also optimize the system structure and using new controller for easy maintainability and testability.

Future work requires more effort on noise control and more test on EAST device.

## ACKNOWLEDGMENT

This work is supported by National Key R&D Program of China (Grant No: 017YFE0300500, 2017YFE0300504).